\documentclass[fleqn,12pt,twoside]{article}
\usepackage{espcrc1}
\usepackage{amssymb} 

\newcommand\NPB{{Nucl. Phys.} B}

\newcommand\PLB{{Phys. Lett.} B}
\newcommand\PR{{Phys. Rep.}}

\newcommand\PRL{Phys. Rev. Lett.}
\newcommand\PRC{{Phys. Rev.} C}
\newcommand\PRD{{Phys. Rev.} D}

\newcommand\JPG{{J. Phys.} G}

\newcommand\CQG{Class. Quant. Grav.}

\def\jou#1#2#3#4{{#1} {\bf #2} (#4) #3}

\def\j{\psi} 
\def\l{\lambda}
\def\o{\omega} 
\def\O{\Omega}
\def\aO{\bar \Omega}

\def\BC{{\mathbb C}}
\def\BZ{{\mathbb Z}}

\def\ran{\rangle}
\def\lan{\langle} 

\def\be{\begin{equation}}
\def\ee{\end{equation}} 

\def\eref#1{Eq. (\ref{#1})}

\title{Baryonic coherent state formation from small domain 
       disoriented chiral condensates} 

\author{S.M.H. Wong\address{Department of Physics, The Ohio State University, 
        Columbus, Ohio 43210, U.S.A.}
        \thanks{This work was supported by the U.S. Department of Energy under 
                grant no. DE-FG02-01ER41190.}  
        and 
        J.I. Kapusta\address{School of Physics and Astronomy, University of Minnesota, 
        Minneapolis, MN 55455, U.S.A.}
        \thanks{This work was supported by the U.S. Department of Energy under 
                grant no. DE-FG02-87ER40328.}  
} 
       
\begin{document}

\maketitle

\begin{abstract}
Rare hyperon yields such as the $\O$ and $\bar \O$ in heavy ion 
collision experiments are hard to be reproduced by numerical models. 
This, in combination with the thermal fit to SPS data, seems to call for 
a new production mechanism beyond the usual ones. Small domain disoriented 
chiral condensates (DCC) were proposed to be such a source of rare hyperons 
through skyrmion formation at the chiral phase transition. Here skyrmions
are treated as coherent states of baryons on a compact manifold so that
the distribution of baryons produced from a skyrmion can be known.
From this more refined treatment, the number of topological defects produced 
are more than doubled to 30 or more and the domain size at the SPS is found 
to be even smaller than before at 1.1--1.5 fm. {\it It is imperative therefore
not to use only pion distribution but other means for observing DCC.}
\end{abstract}

\section{INTRODUCTION} 

Ever since the first proposal \cite{kmr} of using strange hadrons as a 
probe of the formation of the quark-gluon plasma (QGP) in relativistic heavy 
ion collisions, the yields of strange hyperons remain a large part of the 
standard observables for verifying the detection of any enhancement. In this 
respect, it is also a goal of numerical models to simulate the amount of this  
yield in order to make sure that we have understood the underlying production 
mechanisms that may occur in these collisions. Data from the 
Super Proton Synchrotron (SPS), which came out gradually over a number of 
years, have allowed theorists to perform exactly this kind of comparison. 
It turned out that the $\O$ and $\aO$ had proved to be the most stubborn
and challenging of the hyperons for numerical models. At the same time thermal
fits can be done to most yields of the hadrons without too much of a problem 
with the exception of the $\O$ and $\aO$. In \cite{rl} it was shown that the 
fits improved dramatically by an order of magnitude as measured by the value 
of $\chi^2$ (one can of course fit even the $\Omega$ by lifting the 
requirement of the best fits should have small $\chi^2$ and allow for 
arbitrary large $\chi^2$). The general tendency is that there are more $\O$ and 
$\aO$ than expected even after taking into account of QGP formation (in which 
case thermal model should suffice). This indicates that another production 
mechanism is at work.

\section{PRODUCING HYPERONS FROM DISORIENTED CHIRAL CONDENSATES}

It has been pointed out in \cite{kw} that DCC could produce hyperons. 
Although they were more readily associated with pions, hyperon production 
is indeed possible when topological defects are created during the chiral 
phase transition. The remnants in this case are the so-called skyrmions
from the Skyrme Model. Skyrmions are well known to be baryons, hence the 
connection of hyperons to DCC. To test this possibility against data,  
one has to have two pieces of information. One is the likelihood of a defect
being created in the phase transition, and the other is the relative abundance
to one another of all the observable baryons that will be produced from a 
skyrmion. The first piece of information was obtained in ref. 
\cite{ks}. It tells us that in terms of the correlation length $\xi$, there 
should be about $0.8$ skyrmion and antiskyrmion per correlation volume $\xi^3$. 
By itself, this is not sufficient for verifying the hypothesis of hyperon 
production from DCC. In \cite{kw} one moved forward regardless by taking an 
assumption on the second piece of needed information, that is equal production 
probability of any octet and decuplet baryons from a skyrmion was assumed. 
From this one can deduce from the SPS data that $\xi \sim 2.0$ fm. 
This is within the range expected from theoretical considerations \cite{rw1,gm} 
and is too small for observing DCC using pion distribution \cite{rw1}. 
Therefore the possibility is compatible with the existing data and all available 
information. It is now time for a refined treatment. Improvements can be done 
in a number of ways but shall be restricted to the removal the assumption of 
equal likelihood of any type of baryons being produced from a skyrmion.

\section{BARYON DISTRIBUTION IN A SKYRMION}  

In experiments one cannot observe a skyrmion directly, since it is not a 
single observable hadron. It is essential to know exactly what the skyrmion 
is in terms of the well known baryons. In \cite{me1} it was argued that a 
skyrmion must necessarily be a coherent state of baryons. 
In other words it is a superposition of physical baryon states. 
Being faithful to the original Skyrme model and the set of wavefunctions 
derived from it using the collective coordinate approach \cite{anw}, 
one can construct coherent states using these as ingredients. The baryon 
wavefunctions of \cite{anw} all live on $S^3$. Therefore the coherent 
states must also live in this compact space. Coherent states on compact 
manifolds may sound unfamiliar to a heavy ion physicist but not so in 
the field of quantum gravity and functional analysis. The basic method 
and related issues are outlined in a series of papers by different 
authors \cite{ha,tt0,tt1,tt2}. 
The method itself is not based explicitly on observable states of the  
Hamiltonian but rather on the position space of the manifold itself. 
It so happens that the collective coordinate quantization of the Skyrme 
model produces both spin one-half physical states and integral spin 
nonphysical states \cite{anw,fr}. Therefore one has to be careful that 
no nonphysical states are actually present in the coherent states that 
represent the skyrmions. 

The method of \cite{ha,tt0} requires the construction of annihilation 
operators and from there one obtains the coherent states as the simultaneous 
eigenstates of the operators. The annihilation operators in the case of the 
Skyrme model are of the form 
\be \hat A_b = \exp \left (-\frac{1}{2\l\,\o} \hat J^2\right ) \, \hat a_b \,
               \exp \left ( \frac{1}{2\l\,\o} \hat J^2\right )          
\label{eq:a-op-s3}
\ee
where $\lambda$ is a constant in the Skyrme Hamiltonian, $\omega$ is a 
fundamental energy scale of the problem, $\hat a_b$ ($b=0,1,2,3$) are the 
position operators on the compact manifold of $S^3$ and $\hat J_i$ are the spin 
angular momentum operators. One can easily see that the eigenstates must be 
given by 
\be |\j,a \ran = \exp \left (-\frac{1}{2\l \o} \hat J^2\right ) |a \ran 
\label{eq:cs-s3} 
\ee 
with the eigenvalues given by the position label $a$. This is, however, 
not sufficient because the real $a$ does not contain enough information 
to label a coherent state. It must carry information of both position and
momentum. To qualify it has first to be analytically continued 
to complex $a^\BC$. One now has  
\be |\j,a^\BC \ran = \exp \left (-\frac{1}{2\l \o} \hat J^2 \right ) |a^\BC \ran   \;. 
\ee 
To check that it has the required physical properties to be a coherent state
of baryons, a complete set of spin and isospin states of the Skyrme 
Hamiltonian is inserted 
\be |\j,a^\BC \ran = \sum_{j,m,n \in \BZ,\BZ/2} \exp \left (-\frac{1}{2\l \o} 
                     \hat J^2\right ) |j,m,n \ran \lan j,m,n |a^\BC \ran   \;. 
\label{eq:cs-s3-c} 
\ee 
As already mentioned the sum over states includes both integral and half-integral 
(iso)spin states which naturally presents a problem. The simplest solution seems 
to be to drop all the integral spin states from \eref{eq:cs-s3-c} but that 
apparently does not solve the problem. The Skyrme model is usually quantized 
using the $SU(2)$ collective coordinates \cite{anw}, hence the coordinates  
operators are $\hat a_b$. The algebra of $SU(2)$ dictates that for example 
$[\hat J_3, \hat a] \sim [\hat I_3, \hat a] \sim \mbox{$1 \over 2$} \hat a $,  
so $\hat a$ acting on $|j,m,n\ran$ will change $(j,m,n)$ from half-integral
numbers to integral numbers. Therefore a pure fermion version of 
\eref{eq:cs-s3-c} cannot stay within the fermionic space forever. 
Physical and unphysical states will mix. 

A solution to this is to first map from $SU(2)$ to $SO(3)$ \cite{me1,bmss} 
using $A \tau_i A^\dagger = \tau_j R_{ji}$ where $A = a_0+i \tau_i a_i$.  
The resulting nine $SO(3)$ coordinate operators $\hat R_{ij}$ keeps 
$|j,m,n\rangle$ a fermion state if it is one to begin with, because of 
the algebra 
$[\hat J_3, \hat R] \sim [\hat I_3, \hat R] \sim \hat R  \;.$ 
The solution is then to use the $SO(3)$ operators in combination
with the original $SU(2)$ baryon states of \cite{anw}. The annihilation 
operators become
\be \hat A_{ij} = \exp \left (-\frac{1}{2\l\,\o} \hat J^2\right ) \, \hat R_{ij} \,
                  \exp \left ( \frac{1}{2\l\,\o} \hat J^2\right ) \;.          
\label{eq:a-op-so3}
\ee
So finally one can equate  
\be |\j,a^\BC \ran = \sum_{j,m,n \in \BZ/2} \exp \left (-\frac{1}{2\l \o} 
                     \hat J^2\right ) |j,m,n \ran \lan j,m,n |a^\BC \ran   
              \equiv | S \ran  
\label{eq:cs-s3-c-hz} 
\ee 
with a skyrmion. The probability of the skyrmion being in any of the baryon 
states $|j,m,n\ran$ is given by the modulus squared of the coefficient 
divided by the normalization. 

Clearly that the assumption made in \cite{kw} of equal probability 
for any of the baryon states to emerge from a skyrmion isn't 
exactly correct. Applying the above result at low energy appropriate
to DCC, using the octet and decuplet baryon mass difference and setting 
the fundamental energy scale at $\omega \sim T_c$ since this is the 
only scale available, one finds that the instead of 1:1, the octet 
to decuplet baryon ratio is between 5:1 to 14:1 depending on the 
exact value of $T_c$ used (150--200 MeV). Now proceeding again as 
in \cite{kw} but with the use of this new found information, one deduces 
that the domain size is now even smaller but not unreasonably small 
at $\xi \sim$ 1.1--1.5 fm. Because both the SPS data and the probability of 
skyrmion formation per unit correlation volume do not change, there must be 
more skyrmions to compensate for this reduction in domain size. This increase
is at least double that from the previous estimate of 14 to 30--65.  

In conclusion even with the refined treatment of the skyrmions, 
this mechanism for the production of excess hyperons is still 
compatible with all the existing SPS data and the theoretical expectation
on the small size of the DCC domains. In \cite{rw1} it was mentioned that 
if the system was in equilibrium the domain size would be too small for the 
observation of DCC through pions. This is assuming that one is using pion
distribution as the basis of observation. The mechanism discussed here does
not rely on pions and the small size of the domains will not preclude them
from being observed. In fact they thrives on small $\xi$ \cite{kw,ks}. 
Details of the work done here will be reported elsewhere \cite{me2}. 
Another way to observe DCC also discussed in this conference \cite{sg}, 
when the domain size is small, is by using pion/kaon fluctuations proposed 
in \cite{gk}. Although any attempts so far at detecting DCC have been 
unsuccessful, they are all based on the premise that the domain size has 
to be large. It is definitely about time that this old strategy be set 
aside and new ones be further developed to take into account of the 
possibility that DCC are formed only in small domains.

\end{document}